\documentclass[conference, 10pt]{IEEEtran}
\IEEEoverridecommandlockouts
\usepackage{cite}
\usepackage{amsmath,amssymb,amsfonts}
\usepackage{algorithm}
\usepackage{algpseudocode}

\usepackage{graphicx}
\usepackage{textcomp}
\usepackage{xcolor}
\usepackage{soul}
\usepackage{booktabs}
\usepackage[font=small,skip=5pt]{caption}
\newcommand{\red}[1]{\textcolor{red}{#1}}
\newcommand{\anish}[1]{\textcolor{blue}{#1}}
\newcommand{\nirmal}[1]{\textcolor{magenta}{#1}}
\newcommand{\josh}[1]{\textcolor{orange}{#1}}

\usepackage{verbatim}
\usepackage[scaled=1, varl]{inconsolata}

\usepackage{comment}

\usepackage{array}

\usepackage{listings}
\usepackage{xcolor}

\colorlet{punct}{red!60!black}
\definecolor{background}{HTML}{EEEEEE}
\definecolor{delim}{RGB}{20,105,176}
\colorlet{numb}{magenta!60!black}

\lstdefinelanguage{json}{
    basicstyle=\normalfont\ttfamily,
    stepnumber=1,
    numbersep=4pt,
    showstringspaces=false,
    breaklines=true,
    backgroundcolor=\color{background},
    literate=
     *{0}{{{\color{numb}0}}}{1}
      {1}{{{\color{numb}1}}}{1}
      {2}{{{\color{numb}2}}}{1}
      {3}{{{\color{numb}3}}}{1}
      {4}{{{\color{numb}4}}}{1}
      {5}{{{\color{numb}5}}}{1}
      {6}{{{\color{numb}6}}}{1}
      {7}{{{\color{numb}7}}}{1}
      {8}{{{\color{numb}8}}}{1}
      {9}{{{\color{numb}9}}}{1}
      {:}{{{\color{punct}{:}}}}{1}
      {,}{{{\color{punct}{,}}}}{1}
      {\{}{{{\color{delim}{\{}}}}{1}
      {\}}{{{\color{delim}{\}}}}}{1}
      {[}{{{\color{delim}{[}}}}{1}
      {]}{{{\color{delim}{]}}}}{1},
}

\def\BibTeX{{\rm B\kern-.05em{\sc i\kern-.025em b}\kern-.08em
    T\kern-.1667em\lower.7ex\hbox{E}\kern-.125emX}}
\begin{document}

\title{User-Space Emulation Framework for Domain-Specific SoC Design\\
\thanks{\footnotesize{
 This material is based on research sponsored by Air Force Research Laboratory (AFRL) and Defense Advanced Research Projects Agency (DARPA) under agreement number FA8650-18-2-7860. The U.S. Government is authorized to reproduce and distribute reprints for Governmental purposes notwithstanding any copyright notation thereon. The views and conclusion contained herein are those of the authors and should not be interpreted as necessarily representing the official policies or endorsements, either expressed or implied, of Air Force Research Laboratory (AFRL) and Defence Advanced Research Projects Agency (DARPA) or the U.S. Government.
 }}
}

\makeatletter
\newcommand{\linebreakand}{%
  \end{@IEEEauthorhalign}
  \hfill\mbox{}\par
  \mbox{}\hfill\begin{@IEEEauthorhalign}
}

\author{\IEEEauthorblockN{Joshua Mack}
\IEEEauthorblockA{\textit{ECE}, \textit{The University of Arizona}\\
Tucson, AZ, USA \\
jmack2545@email.arizona.edu}
\and
\IEEEauthorblockN{Nirmal Kumbhare}
\IEEEauthorblockA{\textit{ECE}, \textit{The University of Arizona}\\
Tucson, AZ, USA \\
nirmalk@email.arizona.edu}
\and
\IEEEauthorblockN{Anish NK}
\IEEEauthorblockA{\textit{ECEE}, \textit{Arizona State University}\\
Tempe, AZ, USA \\
anishnk@asu.edu}
\and
\linebreakand 
\IEEEauthorblockN{Umit Y. Ogras}
\IEEEauthorblockA{\textit{ECEE}, \textit{Arizona State University}\\
Tempe, AZ, USA \\
umit@asu.edu}
\and
\IEEEauthorblockN{Ali Akoglu}
\IEEEauthorblockA{\textit{ECE}, \textit{The University of Arizona}\\
Tucson, AZ, USA \\
akoglu@email.arizona.edu}

}

\maketitle
\begin{abstract}
In this work, we propose a portable, Linux-based emulation framework to provide an ecosystem for hardware-software co-design of Domain-specific SoCs (DSSoCs) and enable their rapid evaluation during the pre-silicon design phase. 
This framework holistically targets three key challenges of DSSoC design: accelerator integration, resource management, and application development.
We address these challenges via a flexible and lightweight user-space runtime environment that enables easy integration of new accelerators, scheduling heuristics, and user applications, and we illustrate the utility of each through various case studies.
With signal processing (WiFi and RADAR) as the target domain, we use our framework to evaluate the performance of various dynamic workloads on hypothetical DSSoC hardware configurations 
composed of mixtures of CPU cores and FFT accelerators using a Zynq$^{\textregistered}$ UltraScale+\textsuperscript{{\tiny TM}} MPSoC.
We show the portability of this framework by conducting a similar study on an Odroid platform composed of big.LITTLE ARM clusters. 
Finally, we introduce a prototype compilation toolchain that enables automatic mapping of unlabeled C code to DSSoC platforms.
Taken together, this environment offers a unique ecosystem to rapidly perform functional verification and obtain performance and utilization estimates that help accelerate convergence towards a final DSSoC design.

\end{abstract}

\begin{IEEEkeywords}
DSSoC, Domain-Specific, System on Chip, SoC, Domain-Specific SoC, emulation, automatic application mapping
\end{IEEEkeywords}

\vspace{-5mm}
\section{Introduction}
\label{sec:introduction}

As technology scaling becomes a challenge, System-on-Chip (SoC) architects are exploring the capabilities of Domain-Specific SoCs (DSSoCs) to effectively balance performance and flexibility~\cite{magarshack2003system}.
DSSoC architectures are characterized by a heterogeneous collection of general-purpose cores and programmable accelerators tailored to a particular application domain. 
The uniqueness of DSSoC architectures gives rise to a number of challenges~\cite{sadowski2014design}.

First, the design and implementation of hardware accelerators is time-consuming and complex~\cite{koeplinger2016automatic}.
DSSoCs are characterized by application domains with recurring compute- and/or energy-intensive routines, and an effective DSSoC will require a collection of accelerators built specifically to handle these.
Hardware implementation and functional verification of custom accelerators while meeting area, timing, and power constraints at the system-level remains a significant challenge.

Second, DSSoCs commonly operate in real-time environments where time-constrained applications arrive dynamically.
For a fixed collection of heterogeneous accelerators, this requires dynamic and low-overhead scheduling strategies to enable effective runtime management and task partitioning across these accelerators.
A common approach in enabling rich scheduling algorithms that maximize PE utilization is to model applications as directed acyclic graphs (DAGs).
Assuming DAG-based applications, the complexity of managing a large collection of task-dependencies and prioritizing execution across a variety of custom and general-purpose PEs makes scheduling a non-trivial problem in DSSoCs.

Third, like any heterogeneous platform, it is crucial to provide productive toolchains by which application developers can port their applications to DSSoCs~\cite{uhrie2019machine}.
In particular, target applications must be analyzed in terms of their phases of execution, and the portions of each application that are amenable to heterogeneous execution must be mapped as such to the various resources present on a given DSSoC.
Providing application developers a rich environment by which they can explore different application partitioning strategies contextualized by realistic scheduler models and accelerator interfaces is critical in enabling efficient execution on production hardware.

Finally, in a production DSSoC, effective on-chip communication is crucial to exploit maximum performance with minimum latency and energy consumption.
Hence, there is a need for efficient Network-on-Chip (NoC) fabric that is tailored for a given DSSoC's collection of accelerators.
Together with the aforementioned challenges, it is a complex task to design and evaluate DSSoC architectures.

In this work, we propose an open-source, portable user-space emulation framework \cite{DSSoC_emul} that seeks to address the first three challenges of accelerator design, resource management, and application development in the early, pre-silicon stages of DSSoC development.
This framework is a lightweight Linux application that is designed to be suitable for emulating DSSoCs on various commercial off-the-shelf (COTS) heterogeneous SoCs.
For the above three challenges, it provides distinct plug-and-play integration points where developers can individually integrate and evaluate their applications, schedulers, and accelerator IPs in a realistic and holistic system before a full virtual platform or platform silicon is made available.
Notably, to enable rapid application integration, our framework also includes a prototype compilation toolchain that allows users to map monolithic, unlabeled C applications to DAG-based applications as an alternative to requiring hand crafted, custom integration for each application in a domain.
On top of enabling functional verification for each of these aspects of a DSSoC separately, this unified environment assists in deriving relative performance estimates among different combinations of applications, scheduling algorithms, and DSSoC hardware configurations.
These estimates are expected to assist DSSoC developers narrow their configuration space prior to performing in depth, cycle-accurate simulations of a complete system and accelerate convergence to a final DSSoC design.

The rest of the paper is organized as follows. 
In Section \ref{sec:emulation_framework}, we introduce our proposed framework and describe the functionality of its key components. 
We also describe the interfaces required to integrate new schedulers, applications and processing elements. 
In Section \ref{sec:case_studies}, we present various use-cases of our emulation framework based on real applications from the signal processing domain on COTS platforms. 
In Section \ref{sec:lit_survey}, we review relevant work in literature and conclude in Section \ref{sec:conclusion}.

\section{Emulation Framework}
\label{sec:emulation_framework}
\begin{figure}[t]
\centering
\includegraphics[width=0.85\linewidth]{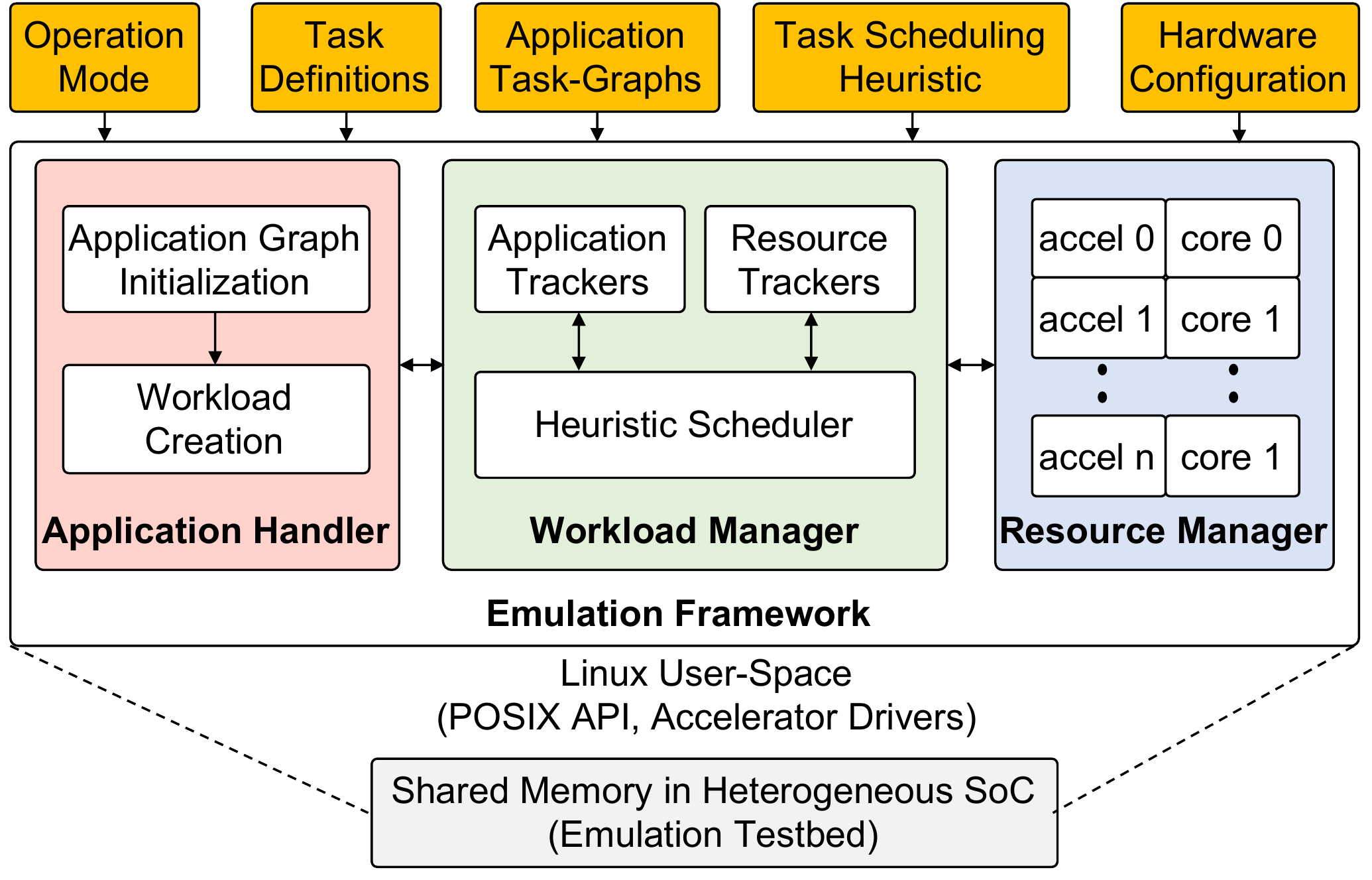}
\caption{Overall architecture of the emulation framework showing the  application handler, workload manager, and resource manager.}
\label{fig:sch_framework_comp}
\vspace{-6mm}
\end{figure}

\subsection{Overview}
Figure \ref{fig:sch_framework_comp} presents a simplified block diagram of our emulation framework. 
It is composed of three key components, (a) the application handler, (b) the workload manager, and (c) the resource manager. 
The application handler is responsible for initializing the framework-compatible representations of all the application task-graphs and create a workload for the framework. 
The workload manager schedules tasks from the DAGs onto the PEs based on the scheduling policy chosen by the user.
The resource manager is used to create the test hardware configuration using the PEs in the SoC and coordinate the execution of the tasks with the workload manager.
The framework uses one of the CPU cores among the available pool of PEs to act as a management processor. This core is dedicated to run the application handler and the workload manager modules. 
The rest of the PEs form the resource pool from which resource manager can instantiate different test hardware configurations. 
All the components of the framework and the tasks for each application are written using C/C++. 
The framework operates in the Linux user-space and requires POSIX thread library~\cite{mueller1993library}. 
This makes it portable across wide range of commercial SoC platforms. 
By default, the framework is integrated with the applications from the signal processing domain, such as Radar and WiFi, to aid the development of DSSoC for software defined radios (SDR).

At the start of an emulation, the framework performs an initialization phase in which the application handler initializes a queue containing the required workload, and allocates the memory required by the emulation workload in the main memory. 
In the same phase, the resource manager initializes the target DSSoC configuration by using the real PEs in the underlying SoC. 
Post the initialization phase, the workload manager drives the emulation by dynamically injecting the applications from the workload queue and coordinating with the resource manager to schedule tasks on the idle PEs.
Before termination, the framework collects the scheduling statistics for all the applications and their tasks.
These statistics can later be used to evaluate the performance of the emulated DSSoC. 
The communication between different PEs is performed using the 
shared memory 
of
the platform.
As a result, while this framework can assist in hardware, scheduler, and application design, it currently is limited in its ability to handle hypothetical NoC architectures.
In the subsequent subsections, we present details of all the components in our framework, 
and we detail the steps that must be taken to integrate new features.

\begin{figure}[t]
    \centering
    \includegraphics[width=0.9\linewidth]{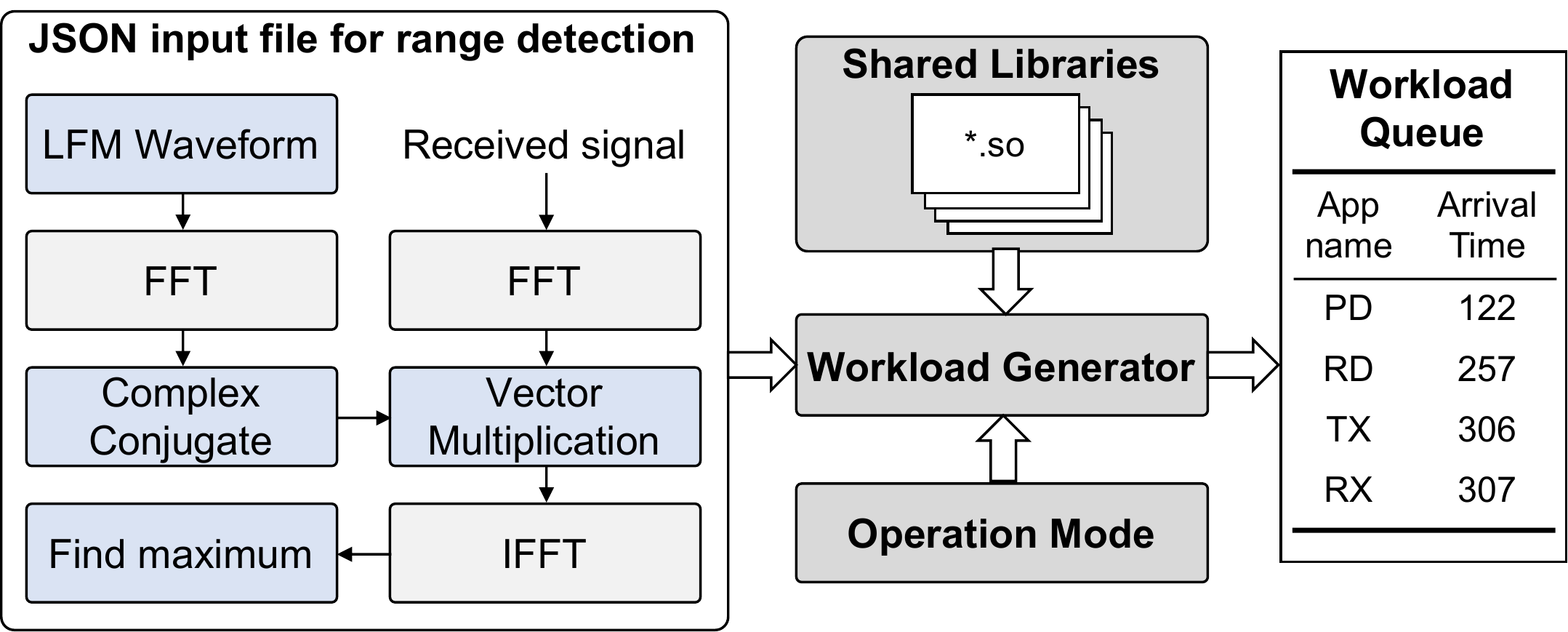}
    \caption{Block diagram describing the application handler}
    \label{fig:app_handler}
    \vspace{-4mm}
\end{figure}

\subsection{Application Handler}


In the framework, the application handler is responsible for parsing and initializing the applications from their respective task-graph representations. 
Figure~\ref{fig:app_handler} presents the functionality of the application handler.
Each user application in our framework consists of two components: a shared object file that contains the functions (\emph{kernels}) that a user's application requires, and a JSON-based DAG that describes their dependency relationships.
These JSON-based DAGs describe the kernels in a given application along with their interconnections, communication costs (data transfer volumes), execution time cost on supported platforms (CPU, accelerator), and the names of the function symbols associated with each kernel in the user's shared object application. 
We will take radar-based \emph{range detection} as our motivating application in the domain of software defined radio. 
The task flow graph for range detection is shown as an input to the workload generator in Figure \ref{fig:app_handler}. 
Its JSON representation is shown in Listing~\ref{lst:range_detection}.

\begin{lstlisting}[language=json,firstnumber=1,basicstyle=\scriptsize\ttfamily,label=lst:range_detection,caption=Range Detection JSON,float=t]
"AppName": "range_detection",
"SharedObject": "range_detection.so",
"Variables": {
    "n_samples": {
        "bytes": 4,
        "is_ptr": false,
        "ptr_alloc_bytes": 0,
        "val": [0, 1, 0, 0]
    },
    "lfm_waveform": {
        "bytes": 8,
        "is_ptr": true,
        "ptr_alloc_bytes": 2048,
        "val": []
    },
    "rx": {
        "bytes": 8,
        "is_ptr": true,
        "ptr_alloc_bytes": 2048,
        "val": []
    },
    "X1": {
        "bytes": 8,
        "is_ptr": true,
        "ptr_alloc_bytes": 4096,
        "val": []
    }
    ...
},
"DAG": {
  "LFM": {
    "arguments": ["n_samples", "lfm_waveform"]
    "predecessors": [],
    "successors": ["FFT_1"],
    "platforms": 
    [{"name":"cpu", "runfunc":"range_detect_LFM"}]
  },
  "FFT_0": {
    "arguments": ["n_samples", "rx", "X1"]
    "predecessors": [],
    "successors": ["MUL"],
    "platforms": 
    [{"name":"cpu", "runfunc":"range_detect_FFT_0_CPU"},
     {"name":"fft", "runfunc":"range_detect_FFT_0_ACCEL", "shared_object":"fft_accel.so"}]
  },
  ...
  "MAX": {
    "arguments": ["n_samples", "corr", "index", "max_corr", "lag", "sampling_rate"]
    "predecessors": ["IFFT"],
    "successors": [],
    "platforms": 
    [{"name":"cpu", "runfunc":"range_detect_MAX"}]
  }
}
\end{lstlisting}

In Listing~\ref{lst:range_detection}, the \texttt{AppName}, \texttt{SharedObject}, and \texttt{Variables} keys give global information about the application: namely its name within the framework, the shared object that contains the implementation for each function referenced, and the list of all program variables that will be required by nodes within this application.
The \texttt{Variables} key, in particular, has a value that is heavily application dependent and defines the storage requirements and initialization values for any variable in the program.
Each variable is named by its key, and the values inside -- \texttt{bytes}, \texttt{is\_ptr}, \texttt{ptr\_alloc\_bytes}, and \texttt{val} -- refer respectively to the number of bytes it requires to represent its type, whether this type is itself a pointer, the amount of storage that pointer requires, and a list of initial bytes with which to populate this variable.
As an example, the variable \texttt{n\_samples} was originally a 32-bit integer data type with a value of 256.
As such, it is given 4 bytes of storage space, and it is initialized with a little-endian representation of 256 as the byte vector \texttt{[0,1,0,0]}.
As another example, \texttt{lfm\_waveform} was originally a floating point array for 512 32-bit floats, or 2048 bytes.
Therefore, this value is given 8 bytes (as pointer types are 8-bytes on 64-bit systems), it is flagged as a pointer type, and this variable itself is assigned a location in the heap that is allocated for 2048 bytes upon initialization by the framework.

The \texttt{DAG} key in Listing~\ref{lst:range_detection} gives  structure of the application graph itself, with each key corresponding to a node in the application graph containing information about its predecessors,  successors, and supported execution platforms. 
On application startup, the runtime finds the shared object file referenced in the application's JSON, and begins parsing the graph.
As graph parsing proceeds, it looks up every \texttt{runfunc} it finds in the corresponding shared object and associates it with each given DAG node.
Optionally, each ``platform" in a node can include a custom shared object that is referenced specifically to look up that function, such as an FFT invocation that references an ``fft\_accel.so" shared object as shown in the ``FFT\_0" node.
With all applications parsed, the application handler performs initialization of each instance of the requested applications by initializing all of an application's variables as specified in the JSON.
After this, it proceeds to generate the requested workload.
The workload can be generated to run in either \textit{validation} or \textit{performance} mode. 
Validation mode involves generating all application instances and injecting them at t=0, with the emulation finishing once all applications are complete.
Performance mode involves generating a probabilistic trace, where applications are given injection times $t\in [0, t_{end})$ and injected throughout the emulation, with the process finishing once a defined time limit $t_{end}$ is reached. 
In the performance mode, user needs to provide the time period for the injection along with the probability of injection.

As an example, a user may wish to execute three instances of range detection in validation mode. 
Given this request, the framework will parse all available applications, and it will output an error if, at the end of this process, it has not detected range detection as referenced by its \texttt{AppName}.
Assuming the framework was able to find and parse the archetypal instance of range detection, it will then instantiate three copies of this base application. 
Each application instance will have all its variables allocated and initialized as described in the JSON.
After initialization, the application will be enqueued into a workload queue and passed to the \textit{workload manager} to emulate application arrival and scheduling.

To integrate new applications, a developer has three choices. 
First, they can build a DAG based application entirely from scratch, compile it into a shared object of kernels, and link them together with a hand-crafted JSON-based DAG representation. 
Second, they can choose to leverage the existing library of kernels present in other applications and define a new application simply by linking them together in a novel way.
In this way, many application domains can be rapidly implemented through piecemeal combinations of common kernels solely through defining how they become linked together.
Third, a developer can utilize an automated workflow provided as a part of the framework that allows for automatic, if less optimized, conversion from monolithic C code into DAG-based applications.
Further details about the functionality and capabilities of this third option are presented in Section~\ref{subsec:c_to_dag_conversion}.

\subsection{Workload Manager}

The workload manager drives the emulation in the framework. 
It is responsible for tracking the emulation time, injecting applications, implementing scheduling policies, and coordinating with the resource managers to execute the tasks on the PEs.
The workload manager uses the workload queue from the application handler and the task scheduling algorithm from the user as its inputs. 
At run-time, the user is given the option to select either one of the available scheduling policies from the library or use the custom scheduling algorithm. 
The default scheduling library is composed of minimum execution time ($\texttt{MET}$), first ready-first start ($\texttt{FRFS}$), earliest finish time ($\texttt{EFT}$), and random ($\texttt{RANDOM}$). 

\begin{figure}[t]
\centering
\includegraphics[width=0.65\linewidth]{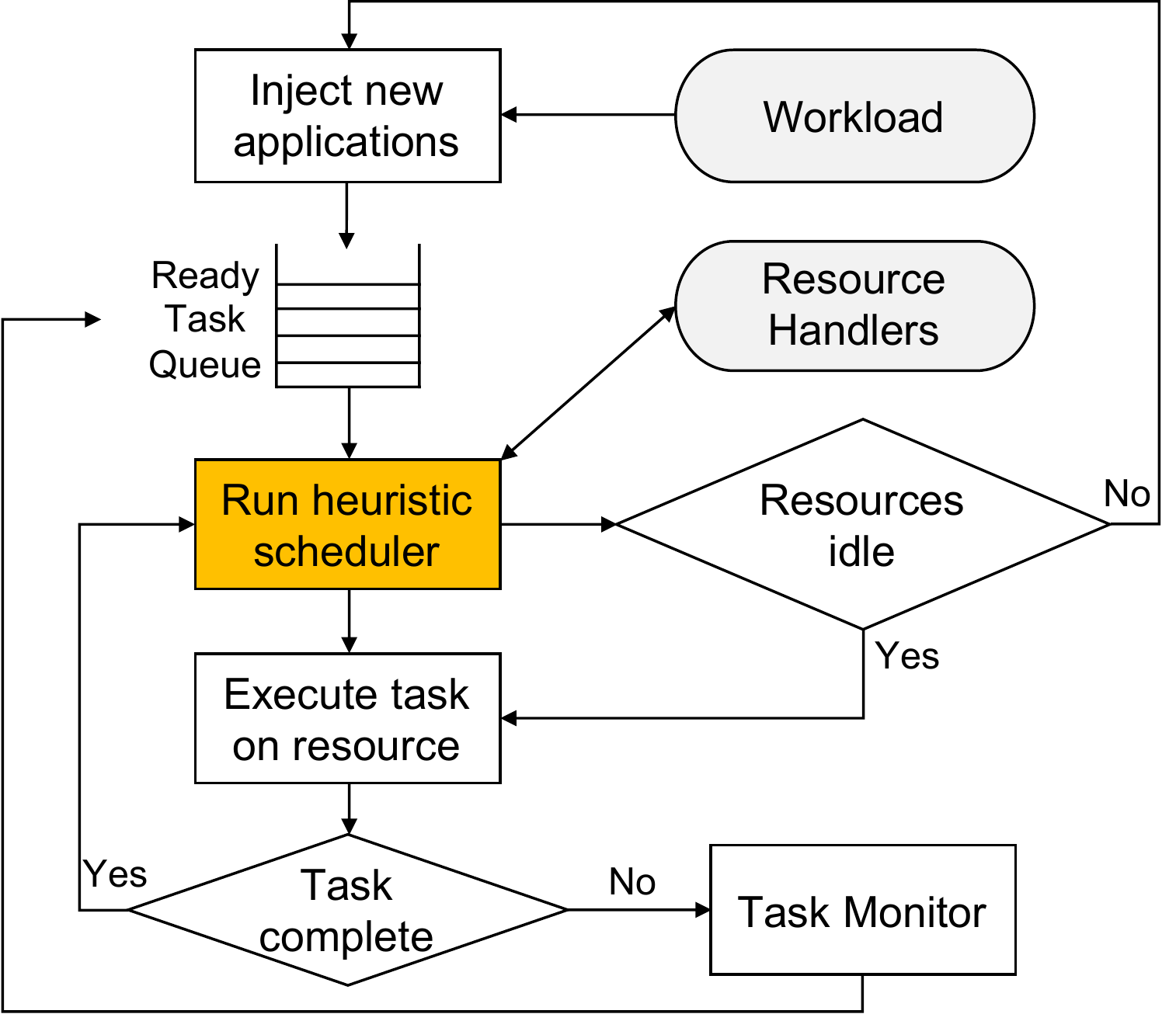}
    \caption{Flowchart on execution of the workload manager}
    \label{fig:workload_manager}
\end{figure}

Figure \ref{fig:workload_manager} presents the basic execution flow of the workload manager in our framework. 
It begins by capturing the system clock as the reference start time for the emulation. 
All the arrival timestamps in the workload queue are relative to this reference start time. 
In this paper, we define emulation time as the time spent in execution after capturing the reference start time.
The workload manager regularly compares the arrival time of an instance at the head of the workload queue with the current emulation time. 
If the current emulation time exceeds the instance arrival time, then it dequeues the head entry (application instance) from the workload queue and injects the instance into ongoing emulation. 
The workload manager appends the head nodes of the newly injected application DAGs into the ready task list. 
The ready task list tracks the tasks that are ready to be executed on the emulated SoC resources. 
After injecting new applications, the workload manager monitors the completion status of the running tasks via resource handler objects. 
A resource handler object is used to manage the communication and synchronization between the workload manager and the resource manager threads. 
Each PE in the emulated SoC is assigned a dedicated resource handler object. 
After monitoring PEs, the workload manager updates the ready task list with the outstanding (unexecuted) tasks. 
An outstanding task is appended in the ready task list if all of its predecessor tasks are completed. 
Next, the user-selected scheduling policy is applied on the ready task list and the tasks selected for scheduling are removed from the list. 
These tasks are communicated to the resource managers of their assigned PEs via resource handlers. 

To utilize a user-defined scheduling policy, an additional policy needs to be defined in \texttt{scheduler.cpp} and a dispatch call needs to be added in the same file's \texttt{performScheduling} function.
This new policy must accept parameters such as the ready queue of tasks and handles for each of the ``resource handler" objects.
Each task consists of a DAG node data structure with all the information necessary for scheduling, dispatch, and measurement of a single node's performance throughout the framework.
Each resource handler object is associated with a unique PE. 
It is composed of fields that track PE availability, type, and id along with its workload and synchronization lock.
The PE availability field is used to communicate resource state
between the workload and resource managers. 
A PE's availability status 
can be $\textit{idle}$, $\textit{run}$, or $\textit{complete}$. 
A thread monitoring or modifying the status field should acquire the PE's synchronization lock, read or write to the status field, and release the lock. 
Integrating a new scheduling algorithm should begin by checking the availability for all the PEs by querying whether their status field indicates they're \textit{idle}. 
Next, the algorithm performs the task to PE mapping on the ready tasks and transfers them over to the resource manager of their mapped PEs via resource handlers. 
Then, the algorithm commands the resource manager to start executing the task by modifying the PE state to $\textit{run}$. 
The resource manager notifies the task completion to the workload manager by modifying the status to $\textit{complete}$. 
Post notification, the workload manager appends the outstanding tasks in the ready list and updates PE status to $\textit{idle}$.
We encourage users to refer to the implementation of the FRFS scheduler in the shared project \cite{DSSoC_emul} to integrate a new algorithm.
\begin{figure}[t]
\centering
\includegraphics[width=0.8\linewidth]{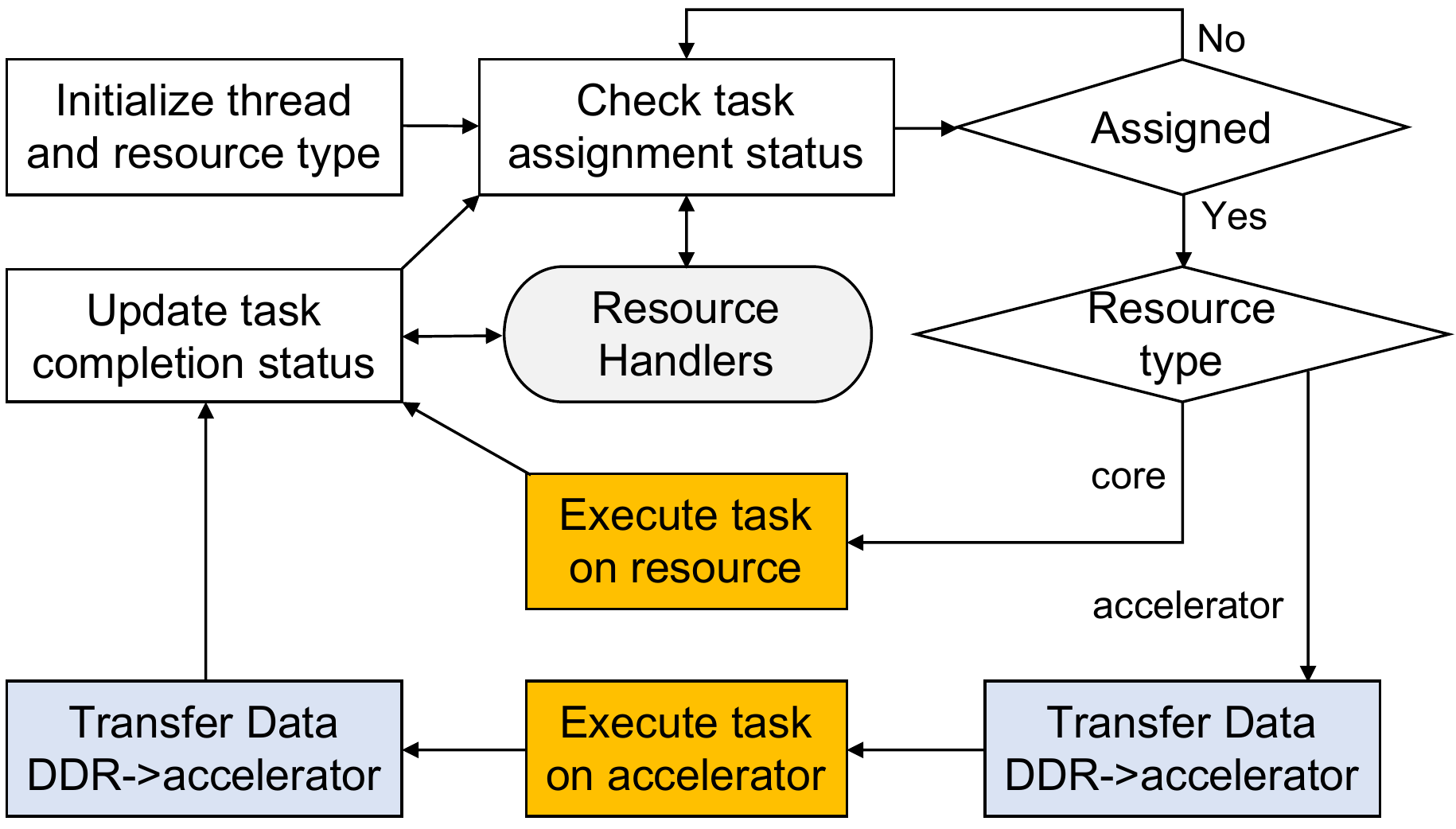}
    \caption{Flowchart on execution of the resource manager}
    \label{fig:resource_manager}
    \vspace{-4mm}
\end{figure}

\subsection{Resource Manager}


At the start of emulation, the framework reads the number and types of PEs from the input configuration file and initializes the dedicated threads of the resource manager for each of the PE. 
These threads are responsible for controlling the operations on their assigned PEs. 
These operations involve the execution of the assigned task, manage the data transfer in between the main memory and the custom accelerator (if required), and coordinate the PE availability status with the workload manager. 
If the input PE type is CPU, then the framework assigns the affinity of its resource manager thread to one of the unused CPU cores in the underlying SoC. 
For all other PE types, 
their resource manager thread assignment begins with the unused CPU cores and then they are evenly distributed among all the CPU cores in the resource pool. 
To derive relative performance estimates, it is recommended to instantiate a test configuration such that all of the resource manger threads are assigned to a separate CPU core to reduce the impact of context switching among the threads.

Figure~\ref{fig:resource_manager} presents a basic block diagram of a resource manager thread. 
It uses the resource handler object to communicate and synchronize with the workload manager. 
After initialization, it checks the task assignment status for the resource in its resource handler. 
If a task is assigned, depending on its resource types (\texttt{core} or \texttt{accelerator}), it follows the execution step as shown in the Figure \ref{fig:resource_manager}. 
If the resource type is \texttt{core}, it executes the executable of the task without any explicit data transfer. 
However, if the resource type is an \texttt{accelerator}, then the resource manager thread transfers the data from the framework memory space (DDR) to the local memory of the accelerator (Block RAM in case of FPGAs), and it follows by commanding accelerator to process the data. 
It monitors the state of the accelerator either using polling or interrupts, and then it transfers data back from the accelerator to the memory space of the emulation framework. 
The framework migrates each accelerator manager thread into sleep state during the processing of the data on the accelerator. 
This allows other manager threads to initiate data transfer and monitor status of their corresponding accelerators if multiple resource managers share the CPU core. 
To integrate new accelerators, a user is expected to implement the blocks required to transfer data between CPU and accelerator, and the programming logic to begin and monitor the completion status of the accelerators. 
In the released repository, we implement DMA interface between accelerators and CPU on ZCU102 platform.

\subsection{Automatic Application Conversion} \label{subsec:c_to_dag_conversion}

\begin{figure}[t]
    \centering
    \includegraphics[width=0.85\linewidth]{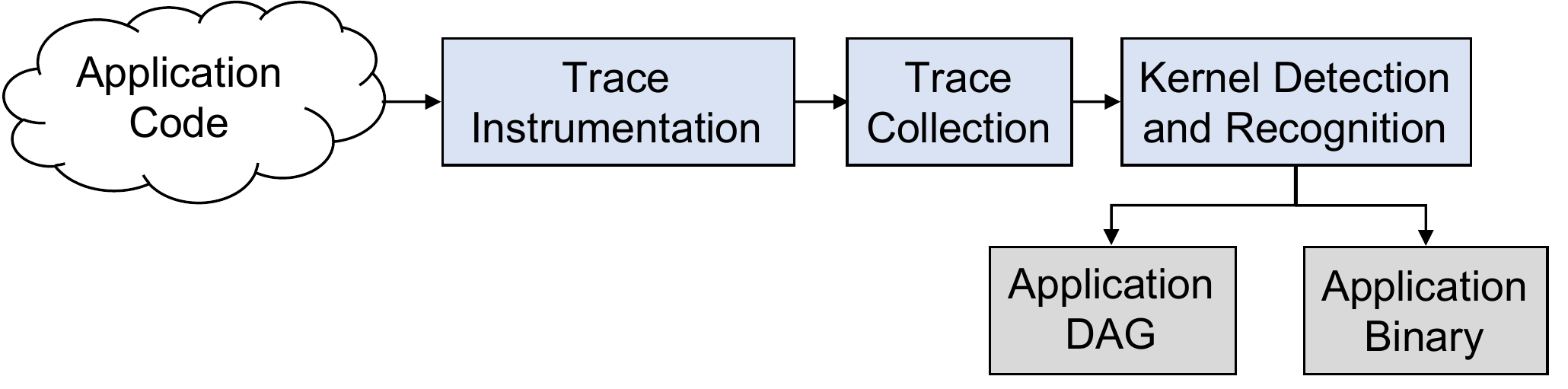}
    \caption{Dynamic tracing-based software flow utilized to automatically convert unlabeled C applications to DAG-based applications}
    \label{fig:automatic_software_flow}
    \vspace{-4mm}
\end{figure}

As an alternative to requiring hand-crafted DAG-based applications, we also provide a basic toolchain that allows for automatic conversion of monolithic, unlabeled C applications to DAG-based applications through a combination of dynamic tracing-based kernel/node detection and LLVM code outlining.
To do this, as shown in Figure~\ref{fig:automatic_software_flow}, we utilize the Clang compiler to convert the application into LLVM~\cite{lattner2004llvm} intermediate representation (IR) and we apply a rich set of tools from the open source LLVM ecosystem. 
Once an application is converted to LLVM IR, we utilize an open source library called TraceAtlas~\cite{traceAtlas} that enables instrumenting standard LLVM code with hooks for dynamic tracing-based analysis.
With the code instrumented, we compile a tracing executable that dumps a runtime trace of its application behavior to disk.
This trace is then analyzed through the TraceAtlas toolchain, and it identifies what sections of the code should be labeled as ``kernels" or ``non-kernels", where a ``kernel" is a set of highly correlated IR-level blocks from the original source code that execute frequently in the base program.
In a broad sense, they are analogous to labeling ``hot" sections in the source program.
With this information, we can partition the original file into alternating groups of ``cold"/``non-kernel" code and ``hot"/``kernel" code.
We then pass this information through an in-house tool, built on LLVM's CodeExtractor module, that uses the information about these code groups to automatically refactor the LLVM IR into a sequence of function calls, where each function call invokes the proper group of blocks necessary to recreate the original application behavior.
Additionally, this in-house tool analyzes the memory requirements of the original application by identifying both static memory allocation in terms of variable declarations as well as dynamic memory allocation by attempting to statically determine the parameters passed into initial malloc/calloc calls.
With this information, along with the outlined source code via LLVM's CodeExtractor, we are able to automatically generate a JSON-based DAG that is compatible with the runtime framework presented here. 
Thanks to the flexibility present in having each node abstracted as a function call, this JSON-based DAG can actually improve an application's execution by replacing a particular node's \texttt{run\_func} with an optimized invocation that has the same function signature if a particular kernel is able to be recognized.
For example, recognizing a naive for loop-based discrete Fourier transform (DFT) would allow this compilation process to substitute in a call to an FFT library or add support for an FFT accelerator.
By compiling the modified IR source into a shared object, we can use it along with the JSON-based DAG to functionally recreate the user-provided application in our runtime framework.
The end result is unlikely to be as optimized and parallelized at this stage as a hand-crafted DAG, but it provides a quick path for porting functionally correct code into the runtime presented.



\section{Case Studies}
\label{sec:case_studies}

\subsection{Overview}
In this section, we present four case studies to demonstrate the usability and portability of our proposed framework. 
In the first study, we use the validation mode of our framework to identify a suitable DSSoC configuration to meet the performance requirements. 
In the second study, we use the performance mode to narrow down on the scheduling policy for a given application domain. 
We demonstrate the portability of our framework by conducting a similar study on a different COTS platform in our third case study. 
As a fourth case study, we illustrate our compilation toolchain that maps unlabeled, monolithic code to a DSSoC.
We begin by providing a brief description of the hardware platforms and signal processing applications we use for our studies.

\subsection{Hardware Platforms and Applications}
\label{subsec:ecp_testbed}
\begin{figure}[t]
\centering
\includegraphics[width=0.6\linewidth]{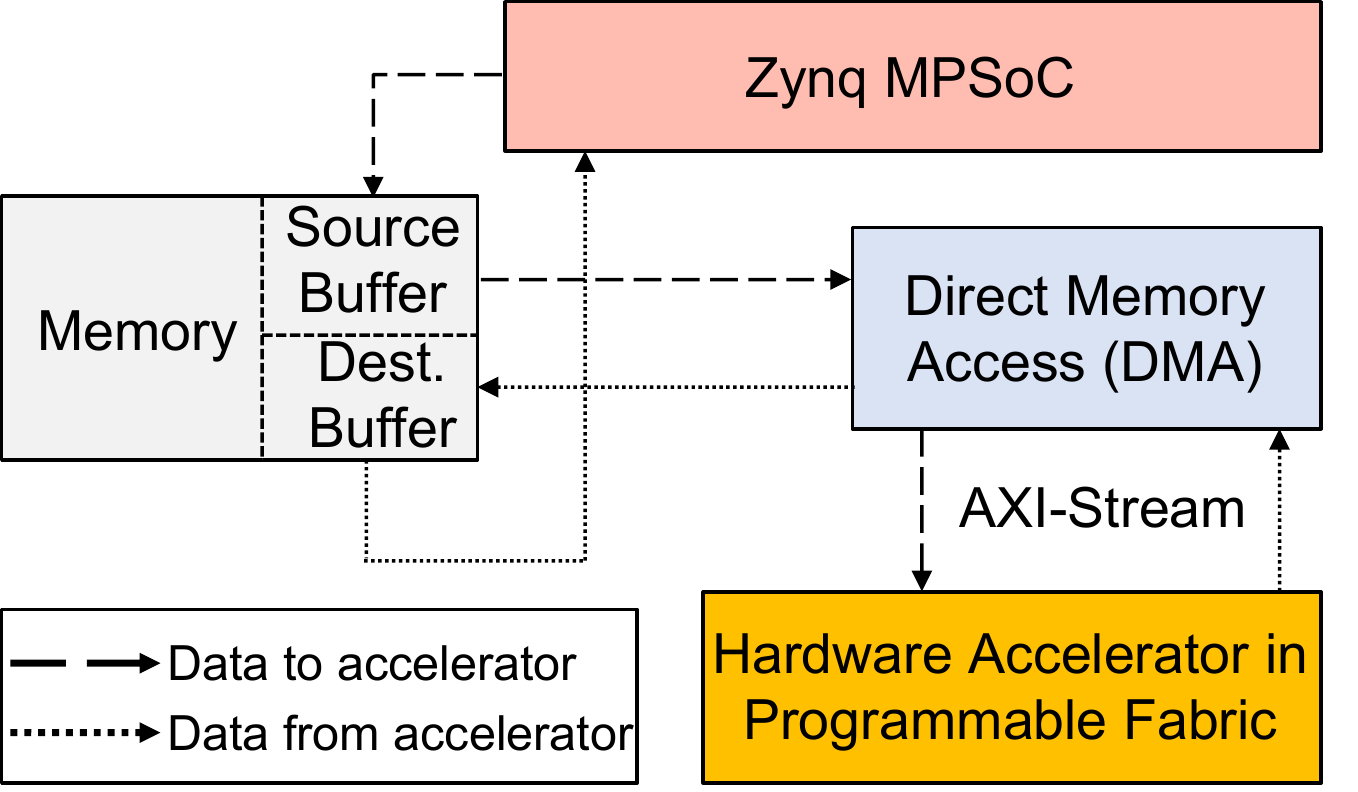}
        \caption{Data flow interface with hardware accelerators}
        \label{fig:accelerator_data_transfer}
        \vspace{-4mm}
\end{figure} 
We use ZCU102 and Odroid XU3 platforms in our case studies. 
ZCU102 is a general-purpose evaluation kit built on top of Zynq$^{\textregistered}$ UltraScale+\textsuperscript{{\tiny TM}} MPSoC \cite{online-ZCU102}. This MPSoC combines general-purpose processing units (quad-core ARM Cortex$^{\textregistered}$ A53 and dual-core Cortex-R5) and programmable fabric on a single chip. 
We create a resource pool composed of two FFT accelerators on the programmable fabric and three general-purpose CPU A53 cores to instantiate different DSSoC configurations. 
We use the fourth A53 core as an overlay processor to run the workload manager and the application handler. 
 
On this platform, we use direct memory access (DMA) blocks to facilitate the transfer of data between memory and hardware accelerators through AXI4-Stream, a streaming protocol~\cite{amba4axi4}. 
The data transfer mechanisms among the host software application, memory and accelerators are illustrated in Figure~\ref{fig:accelerator_data_transfer}. 
We use \emph{udmabuf}~\cite{udmabuf}, an open-source Linux driver that allocates contiguous memory blocks in the kernel space and makes it user-accessible. 
A software application, which operates in the user-space, writes into the shared memory space to transfer data to the programmable logic. 
The DMA IP moves the data to the accelerator for processing and transfers the computed output to the shared memory. 
The software application then reads the data coordinated with the appropriate control logic from DMA and the accelerator. 
\begin{figure}[t]
\centering
\includegraphics[width=0.75\linewidth]{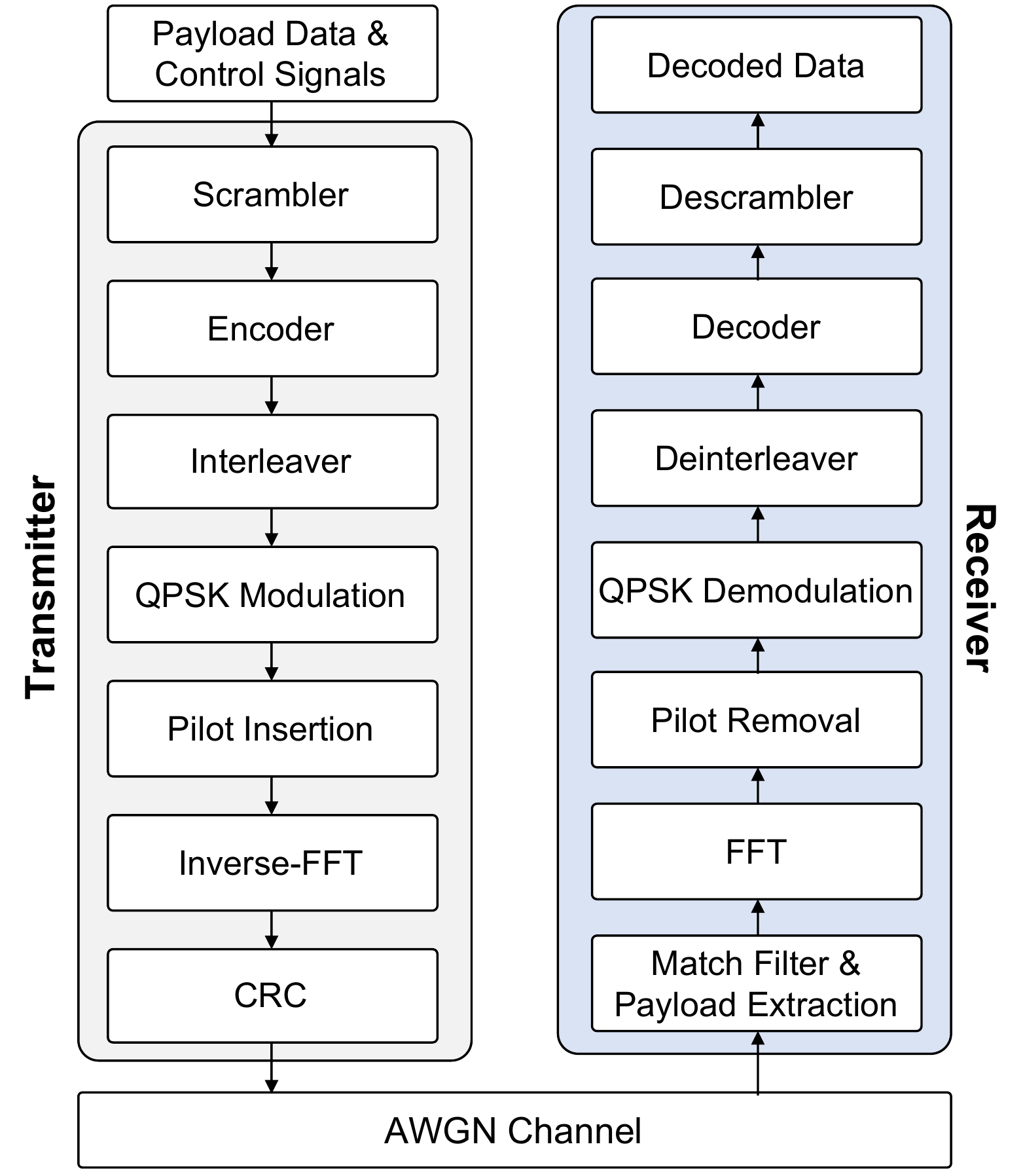}
    \caption{Block Diagram of WiFi Transmitter and Receiver Applications}
    \label{fig:wifi_bd}
    \vspace{-4mm}
\end{figure}

\begin{figure}[t]
	\centering
	\includegraphics[width=0.7\linewidth]{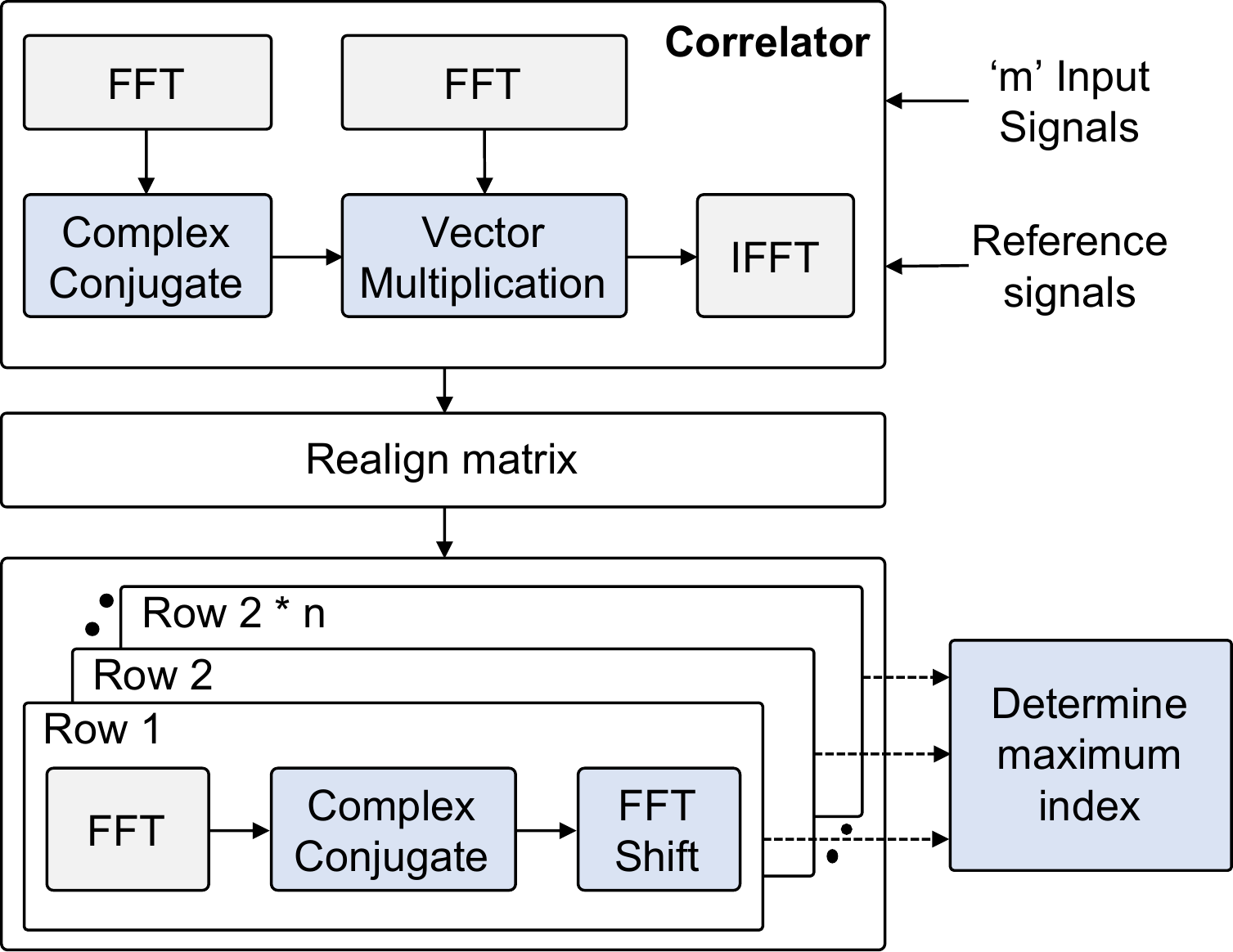}
	\caption{Block diagram of pulse Doppler application where \emph{m} is number of signals and \emph{n} is number of samples in a signal.}
	\label{fig:radar_bd}
	\vspace{-4mm}
\end{figure}


Odroid XU3 is a single board computer, which features an Exynos 5422 SoC. 
The SoC is based on the ARM heterogeneous big.LITTLE architecture \cite{ODROID} in which the LITTLE cores are highly energy-efficient (Cortex-A7) and the big cores (Cortex-A15) are performance-oriented. 
The Cortex-A7 and Cortex-A15 in this SoC are quad-core 32-bit multi-processor cores implementing the ARMv7-A architecture. We use one of the LITTLE core as an overlay processor to run the workload manager and the application handler. 
The remaining four BIG cores and three LITTLE cores form the resource pool to instantiate different DSSoC configurations. 

We select WiFi (RX/TX), Pulse Doppler, and range detection as our representative set of applications in the domain of Software-Defined Radio (SDR).
The WiFi transmitter and receiver applications process 64 bits of data in one frame and are segmented into the kernels shown in Figure~\ref{fig:wifi_bd}.
It is composed of various compute-intensive blocks, such as FFT, modulation, demodulation,  Viterbi decoder, and scrambler \cite{arda2019simulation}. Range detection and Pulse Doppler applications are used in Radar to determine the distance and velocity, respectively, of the target object from the reference signal source. Figures \ref{fig:app_handler} and \ref{fig:radar_bd} present the kernel compositions for the range detection and the Pulse Doppler, respectively. We handcraft the DAG representations for these four applications for our case studies on validation and performance modes.

\begin{figure}[t]
	\centering
	\includegraphics[width=1.0\linewidth]{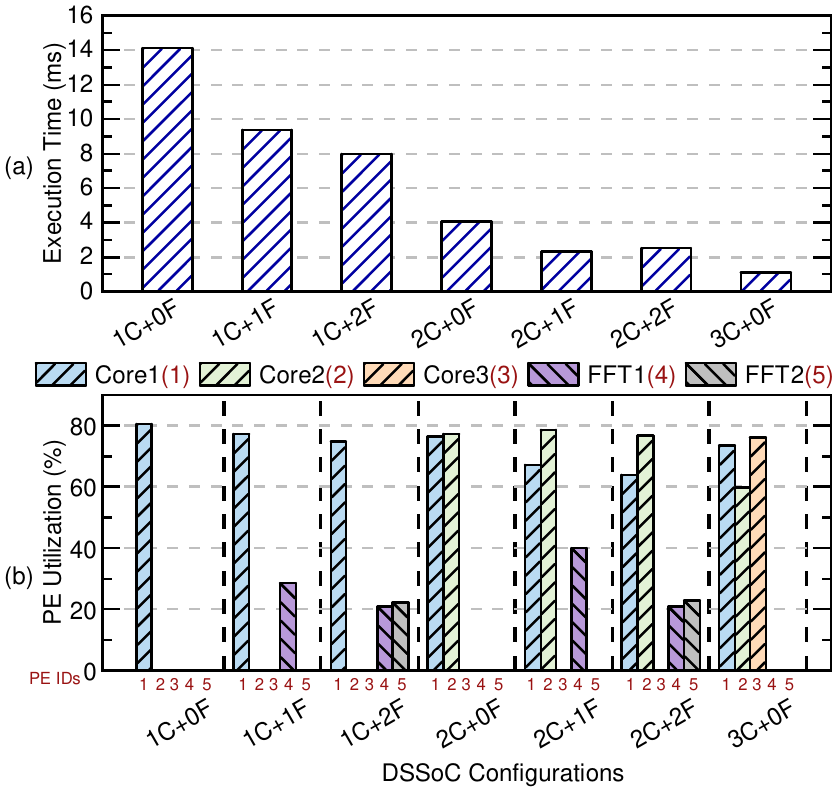}
	\caption{(a) Execution time and (b) PE utilization of a workload on different emulated DSSoC configurations in the validation mode.}
	\label{fig:static_analysis}
	\vspace{-4mm}
\end{figure}

\subsection{Case Study 1: Validation Mode}
The primary use of the validation mode is to functionally verify the integration of an application task-graph, scheduling algorithm, and accelerator in the emulation framework. We also use the validation mode to obtain an estimate on the workload execution time and PE utilization on different SoC configurations. The estimates obtained on the emulation framework are not designed to be cycle-accurate compared to the real silicon chip. Instead, it is designed to assist hardware and software designers to obtain relative performance and PE utilization of a given workload on different target SoC configurations. 
In Figure~\ref{fig:static_analysis}, we plot the execution time (top) and the average PE utilization (bottom) across various DSSoC configurations for a workload composed of single instances of Pulse-Doppler, range detection, and WiFi applications. 
We generate the box plot (top graph) in Figure \ref{fig:static_analysis} based on the execution time for 50 iterations of running this workload. 
We use the ZCU102 platform for this study and dynamically schedule the ready tasks in the given workload based on the FRFS scheduling policy. 
In Figure \ref{fig:static_analysis}, we observe an improvement in the workload execution time with the increase in PE count. However, the increase in CPU core results in a greater improvement in the execution time compared to the FFT accelerators, i.e., execution time improvement is higher as we move from 1Core+1FFT to 2Core+1FFT configuration compared to 1Core+2FFT configuration. We observe this behavior because the input sample count to our FFT  accelerator is only 128. On the ZCU102 platform, an FFT of this size has a faster turn-around time on a CPU core compared to the FFT accelerator. 
The overhead associated with the data transfer from the main memory to programmable fabric and vice-versa in the ZCU102 platform limits the usability of the programmable fabric in processing such a small data set. We observe a negligible difference between the execution times on 2Core+1FFT and 2Core+2FFT configurations. This is because, for the 2Core+2FFT configuration, the resource manager threads for the FFT accelerators share the CPU core. As a result, they keep cyclically preempting each other. The overhead involved in the OS level thread preemption and thread scheduling ends up dominating the benefits of using two FFT accelerators in this configuration. For the remaining configurations in the figure, each resource manager thread executes on a dedicated CPU core. This ensures the improvement in the execution time with the increase in the PEs in the DSSoC configuration.

We calculate PE resource utilization by computing the ratio between the usage time of a PE and the total execution time of the workload. We observe that the utilization of the  CPU cores is significantly higher than the FFT accelerators for the heterogeneous SoC. The maximum CPU core utilization we observe is 80\% for the 1Core+0FFT configuration. Because we regularly execute the scheduling algorithm on the completion of each task, we incur significant scheduling overhead. 
However, in the future, we will incorporate task reservation queues on each PE to reduce the impact of the scheduling overhead. 
From Figure \ref{fig:static_analysis}, we interpret that the 3Core+0FFT configuration has the best execution time. 
If the area and performance are the primary concerns, though, then the 2Core+1FFT configuration is more area efficient while delivering a comparable performance compared to that of the 3Core+0FFT configuration for the given workload.

\begin{table}[t]
    \caption{Application execution time and task count on three core and two FFT accelerators using FRFS scheduling policy}
    \centering
    \begin{tabular}{@{}lcc@{}}
    \toprule
    \textbf{Application} & \multicolumn{1}{l}{\textbf{Execution Time (ms)}} & \multicolumn{1}{l}{\textbf{Task Count}} \\ \midrule
    Range Detection & 0.32 & 6 \\
    Pulse Doppler & 5.60 & 770 \\
    WiFi TX & 0.13 & 7 \\
    WiFi RX & 2.22 & 9 \\ \bottomrule
    \end{tabular}
    \label{tab:app_exe_time_val_mode}
    \vspace{-2mm}
\end{table}

\begin{table}[t]
    \caption{Application instance count used for different injection rates in case study 2}
    \centering
    \begin{tabular}{@{}ccccc@{}}
    \toprule
    \textbf{\begin{tabular}[c]{@{}c@{}}Injection Rate\\ (jobs per msec)\end{tabular}} & \textbf{\begin{tabular}[c]{@{}c@{}}Pulse\\ Doppler\end{tabular}} & \textbf{\begin{tabular}[c]{@{}c@{}}Range\\ Detection\end{tabular}} & \textbf{\begin{tabular}[c]{@{}c@{}}WiFi\\ TX\end{tabular}} & \textbf{\begin{tabular}[c]{@{}c@{}}WiFi\\ RX\end{tabular}} \\ \midrule
    1.71 & 8 & 123 & 20 & 20 \\
    2.28 & 10 & 164 & 27 & 27 \\
    3.42 & 15 & 245 & 41 & 41 \\
    4.57 & 18 & 329 & 55 & 55 \\
    6.92 & 32 & 495 & 82 & 83 \\ \bottomrule
    \end{tabular}
    \label{tab:app_instance_count_perf_mode}
    \vspace{-4mm}
\end{table}

\begin{figure}[]
	\centering
	\includegraphics[width=1.0\linewidth]{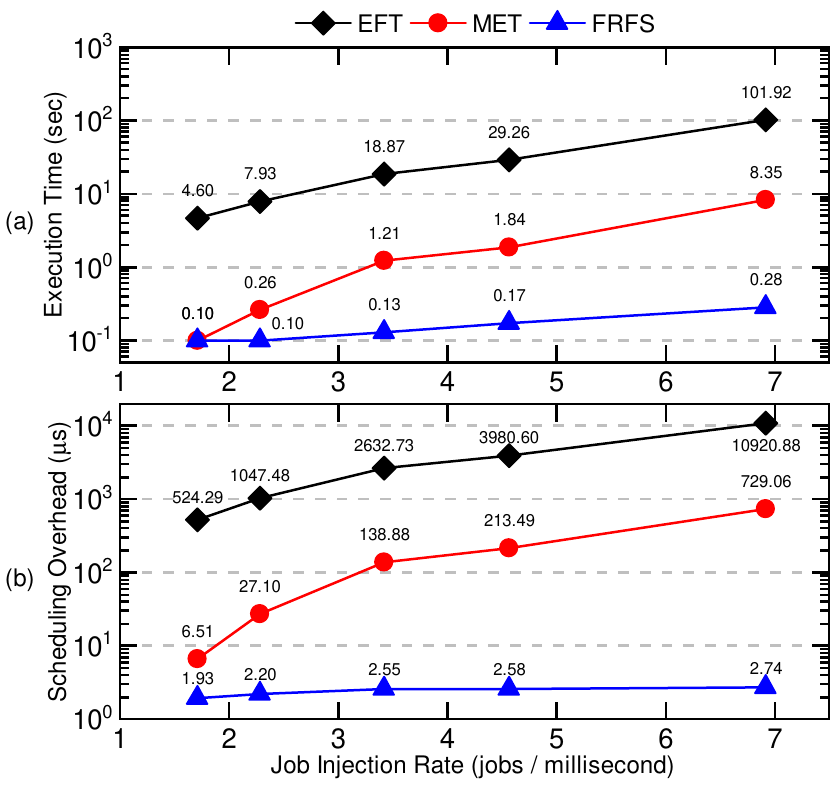}
	\caption{(a) Execution time and (b) average scheduling overhead in performance mode for three schedulers on a 3 core, 2 FFT configuration. The X-axis presents average injection rate in the workload.}
	\label{fig:perf_analysis}
    \vspace{-4mm}
\end{figure}

\subsection{Case Study 2: Performance Mode} \label{subsec:performance_mode}


In this case study, we compare the performance of different scheduling algorithms (FRFS, MET, and EFT) on a DSSoC configuration composed of three cores and two FFT accelerators. 
We operate our emulation framework in the performance mode. This mode is designed to emulate the dynamic injection of the applications on a target DSSoC. 
In the performance mode, the user needs to provide the frequency and probability of injection for each application. 
The user also needs to input the time-frame during which applications are injected. 
For our evaluation, we assume applications are injected periodically with the probability of one in the test time-frame of 100 milliseconds. 
To create a new workload trace, we vary the periodic duration for each application to alter the average injection rate.
Table \ref{tab:app_exe_time_val_mode} presents the standalone execution time for each application on a 3Core+2FFT SoC configuration. Table \ref{tab:app_instance_count_perf_mode} presents the instance count for a given application in each workload trace. Compared to Pulse Doppler, we choose higher injection frequencies for the range detection and WiFi applications because of their shorter execution time and smaller DAG. 

Figure \ref{fig:perf_analysis} presents the workload execution time and the average scheduling overhead for different scheduling policies on a 3Core+2FFT configuration. We calculate scheduling overhead by accumulating the time required to monitor the completion status of the running tasks, update ready queue, run scheduling algorithm on ready tasks, and communicate ready tasks to resource managers for execution. In Figure \ref{fig:perf_analysis}, we observe that the sophisticated scheduling policies, such as EFT and MET, under-perform in terms of workload execution time compared to a simple scheduling policy of FRFS. This is because the computation complexity associated with these schedulers adds up to a significant scheduling overhead as opposed to FRFS policy. The computation complexities for the MET and EFT algorithms are $O(n)$ and $O(n^2)$, respectively. 
Due to the unavailability of the reservation queue on each PE, a scheduling algorithm incurs this overhead every time a task completes its execution on the PE. Eventually, these overheads start accumulating into the workload execution time. In the proposed framework, the complexity of FRFS is equal to the number of PEs in the emulated SoC for the selected group of applications. As a result, we observe a constant scheduling overhead of 2.5 microseconds and a linear increase in the execution time with the increase in the application injection rate for the selected set of applications.

Our framework is successfully able to expose the limitations of underlying design decisions related to the SoC configuration and scheduling policies for a given set of applications. Traditionally, researchers use discrete event-based simulation tools, such as DS3~\cite{arda2019simulation} and SimGrid~\cite{casanova2014versatile}, to develop and evaluate new scheduling algorithms.
These simulators rely on statistical profiling information to realize the performance of general-purpose cores and hardware accelerators. As a result, they are inadequate in capturing scheduling overhead and performing functional validation of the system and IP, as they are designed to operate without real applications and hardware.
Cycle-accurate simulators, such as gem5 \cite{binkert2011gem5,power2014gem5} and PTLSim~\cite{yourst2007ptlsim}, address the drawbacks of discrete event simulators by performing cycle-by-cycle execution of the real applications and scheduling algorithms for the simulated target system or IP.
However, these simulators are slow and primarily used to validate individual IP designs or few specific test-cases for full system validation. 
The turn around time of our emulation framework is substantially lower compared to the cycle-accurate simulators, and its capability to capture the impact of scheduling overheads on the total execution time provide better estimates while performing design space exploration compared to the discrete event simulators.


\begin{figure}[t]
	\centering
	\includegraphics[width=1.0\linewidth]{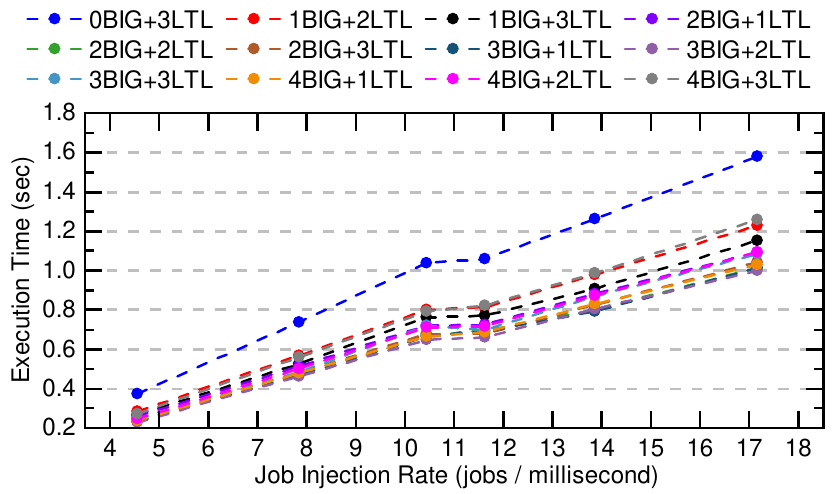}
	\caption{Execution time in different configurations of BIG and LITTLE cores in Odroid XU3 against varying injection rates for a given workload. The framework operates in performance mode and uses FRFS scheduling policy.}
	\label{fig:odroid_perf}
	\vspace{-2mm}
\end{figure}

\subsection{Case Study 3: Performance Analysis on Odroid XU3}
\label{CS3_ODroid}
In this case study, we execute our framework in the performance mode on Odroid XU3 to demonstrate its portability across different COTS platforms. 
In Figure \ref{fig:odroid_perf}, we plot execution time trend with respect to change in job injection rate for different combinations of BIG and LITTLE cores on Odroid XU3. 
To create a test workload, we use an approach similar to the one described in case study 2. For a given injection rate, the same workload is used across all the configurations. We repeat each experiment for multiple iterations and compute the average execution time to plot points in Figure \ref{fig:odroid_perf}. We observe a linear correlation between the workload execution time and the job injection rate. We observe that the configuration composed of three BIG cores and two LITTLE cores, i.e., 3BIG+2LTL, has the best execution time across different job injection rates. We also observe that the configurations 3BIG+1LTL, 4BIG+1LTL, and 2BIG+3LTL perform comparable to the best performing configuration with less than 3\% of impact on the performance. Interestingly, we observe that the workload execution time on the configurations 4BIG+3LTL and 4BIG+2LTL is higher than the execution time on the configuration 4BIG+1LTL. This is because, in the framework, the scheduling complexity of the FRFS algorithm is proportional to the  number of PEs in the emulated SoC. As the PE count in the emulated SoC increases, the scheduling overhead becomes noticeable compared to the task execution time. Furthermore, the lower operating frequency of the overlay processor (LITTLE core) increases the scheduling overhead.



\subsection{Case Study 4: Automatic Application Conversion}
The preceding case studies have primarily focused on exploring performance estimates for different DSSoC configurations and workload scenarios while holding the applications used for evaluation fixed.
However, demonstrating a meaningful path by which application developers can map novel applications to a fixed DSSoC configuration is a similarly critical part of the overall DSSoC design process.
In this case study, we explore the capabilities of our dynamic tracing-based compilation toolchain through automatic mapping of a monolithic range detection C code to our emulation environment.
We target the ZCU102 platform with a configuration composed of 3 cores and 1 FFT accelerator.
As described in Section~\ref{subsec:c_to_dag_conversion}, the toolchain works by using TraceAtlas~\cite{traceAtlas} to dynamically trace the baseline application and extract kernels of interest via analysis of this runtime trace.
In range detection, among the six kernels that are currently detected, three of them consist of heavy file I/O, along with two kernels consisting of two FFTs and one kernel consisting of the IFFT as shown previously in Figure \ref{fig:app_handler}. 
With the kernels identified in this application, we label them as such in the original application LLVM, and we label the remaining contiguous blocks of code as non-kernels.
We then utilize our in-house tool to refactor each contiguous group of kernel/non-kernel LLVM IR into standalone functions and transform the original application into a sequence of function calls, where each outlined function represents one of the nodes in our automatically created DAG.
Together with analysis of the variable and memory requirements for this application, we generate a JSON-based DAG that is able to invoke the outlined functions in an order that preserves the program's correctness.

For this particular application, the two FFTs and one IFFT were implemented as simple for-loop based DFTs and an inverse DFT (IDFT).
As such, to explore the inherent ability to optimize through selecting semantically equivalent but highly optimized \texttt{run\_func} invocations, we compiled an additional shared object library that contained two optimized implementations of the DFT kernel: one that uses FFTW~\cite{Frigo2005FFTW} compiled for ARM to invoke a highly optimized FFT and one that targets the FFT accelerator present on our ZCU102's programmable logic to test our ability to transparently add support for accelerators.
Through hash-based kernel recognition, the \texttt{platform} entries in the DAG JSON were then automatically redirected to this shared object through use of the \texttt{shared\_object} key as first demonstrated in the FFT\_0 node of Listing~\ref{lst:range_detection}.
When replacing this naive DFT kernel with an FFTW call on ARM, including overheads related to FFTW setup and memory allocation, we see a 102X average speedup across both DFT kernel executions, and the application output remains correct.
Similarly, when replacing the DFT kernel with an FPGA-based accelerator call, including data transfer overhead, we see a 94X average speedup across both DFT executions, and the output remains correct.

While these results do require a fairly strict assumption that it is possible to recognize a kernel operationally in an automatic compilation process with no human input, we believe that these results present a promising pathway forward in exploring a generalizable compilation flow for DSSoCs.
Despite this being a first pass implementation of such a compilation flow, we are seeing benefits through new optimization opportunities on the CPU side and through the ability to add automatic support for heterogeneous accelerators without any user intervention or compiler directives.
We are looking forward to enabling further benefits such as support for automatic parallelization of independent kernels via analysis of their runtime memory access patterns and a more generalizable approach for recognizing kernels and pairing them with compatible optimized invocations.






\section{Literature Survey}
\label{sec:lit_survey}
In modern SoC design, pre-silicon verification is the most resource-intensive phase involving functional correctness and constraint checks on the SoC \cite{mishra2017post}.
Considering a three layered approach to SoC design with hardware, resource management and application layers, both, discrete-event \cite{casanova2014versatile,arda2019simulation} and cycle-accurate \cite{binkert2011gem5,power2014gem5,yourst2007ptlsim} based simulation approaches address the challenges involving each layer in isolation. 
As shown in Section~\ref{subsec:performance_mode}, our user-space emulation framework provides an environment where these three layers are considered in tandem.
Compared to a discrete event simulator, our framework allows deriving performance trends among different algorithms rapidly by executing real workloads on a DSSoC that is emulated using COTS hardware. 
By executing on a real hardware platform, the workload turn around time is faster compared to the cycle-accurate simulators in conducting functional validation of the custom IPs and their integration with the applications. This speed-up enables designer to increase the test-coverage in the target DSSoC.

Over the years, commercial FPGA based emulation platforms, such as ZeBu \cite{zebu} and Veloce \cite{veloce}, have been integrated with the SoC development cycle. However, besides their cost, these out-of-the-box platforms require significant effort to enable them for DSSoC development and primarily used for emulating NoC-based designs (including CPU) on the programmable fabric because of their rich debug features. As opposed to these platforms, our framework is designed as a lightweight portable executable to perform design-space exploration and narrow down on probable configurations to target during the early phase of DSSoC development. 

In recent years, researchers have introduced various run-time frameworks to enable the execution of an application in a heterogeneous environment, \cite{augonnet2011starpu, wen2014smart,huang2019cpp,bolchini2018runtime}. These frameworks are targeted for a heterogeneous system in the HPC domain, where a node is composed of CPUs and GPUs. These runtime frameworks provide APIs to design each application in a DAG format and apply scheduling policies on per-application basis. Compared against the earlier implementations, our framework is designed to provide an emulation environment for a DSSoC, where the purpose of a runtime system is not limited to scheduling individual application but it is designed to perform comprehensive design space exploration while iterating over combinations of multiple applications, IPs, and algorithms using COTS platforms. SURF \cite{hsieh2019surf} is one of the most recent runtime framework built for mobile heterogeneous SoC. In SURF, each application DAG is represented as a series of tasks and each task constitutes a set of OpenCL kernels that executes in parallel across heterogeneous PEs. Our framework differs by enabling the complex representation of a DAG. By binding a hardware PE with a software thread, we have fine-grained control on task to PE mapping to execute multiple tasks in parallel as opposed to SURF. Furthermore, our framework provides a preliminary version of a front-end tool that automates the transformation of a monolithic code to DAG. To best of our knowledge, this feature sets us apart from all the previous work in the runtime-framework.

\section{Conclusion}
\label{sec:conclusion}
In this work, we present a portable emulation solution to assist developers in addressing the key challenges of DSSoC development: scheduling algorithms, accelerators, and applications. 
The ability of the proposed framework in evaluating multiple solutions in an unified environment enables designers to derive more accurate performance estimates compared to discreet event simulators and faster turn-around time compared to cycle-accurate simulators. 
We experimentally demonstrated these capabilities using real applications and commercial platforms.
As future work, we will expand our framework to support abstractions like PE-level work queues to enable lower-overhead task dispatch and richer scheduling algorithms.
We will improve our framework's support for device sensors, enabling schedulers to utilize power aware heuristics.
Additionally, we will continue to develop compilation techniques that enable porting monolithic applications to DSSoCs and enabling automatic accelerator support without expert tuning.




\bibliographystyle{IEEEtran}
\bibliography{references}
	
\end{document}